# **Facilitating Human Feedback for GenAI Prompt Optimization**

Jacob SHERSON[1*], Florent VINCHON[2]
[1]Center for Hybrid Intelligence, Dep. of Management, Aarhus University,
* sherson@mgmt.au.dk
[2]Université Paris-Cité and Gustave Eiffel, LAPEA, F-92100

**Keywords:** Generative AI, prompt engineering, human-machine learning, education

Abstract: This study investigates the optimization of Generative AI (GenAI) systems through human feedback, focusing on how varying feedback mechanisms influence the quality of GenAI outputs. We devised a Human-AI training loop where 32 students, divided into two groups, evaluated AI-generated responses based on a single prompt. One group assessed a single output, while the other compared two outputs. Preliminary results from this small-scale experiment suggest that comparative feedback might encourage more nuanced evaluations, highlighting the potential for improved human-AI collaboration in prompt optimization. Future research with larger samples is recommended to validate these findings and further explore effective feedback strategies for GenAI systems.

The incorporation of Generative AI (GenAI) technologies within organizational and research environments emphasizes the urgent need to maximize the performance of these systems to reflect intricate human knowledge and contextual subtleties. Although there is a widespread agreement on the usefulness of GenAI models, optimizing prompts remains a sophisticated challenge that necessitates a thorough examination of the mechanisms that enable productive human-AI interaction.

It is generally essential for a human to be at the center of the improvement loop in order to evaluate the quality of the output produced by a generative AI system like ChatGPT. A human is better equipped to comprehend the context and the circumstances, and can exhibit a refined expertise in the concepts presented, while also adopting a more distinctive and often more pertinent style.

Thus, we recommend the establishment of a human-AI training loop, which begins with a single prompt that is refined and improved over time. In this process, the human plays a pivotal role by evaluating the output and determining its fine-grained properties. By drawing upon their expertise and sensitivity, humans can identify what aspects of the output are more useful and effective. The integration of critical feedback and knowledge of AI-type virtual assistants by experts is crucial to enhancing the quality of Gen AI outputs. This, in turn, leads to a virtuous cycle of continuous skill improvement between AI and humans, ultimately resulting in better acceptance and feedback from users and companies.

However, it is often difficult for domain experts to explicitly specify their knowledge within their field of expertise. Indeed, expressing a nuanced and contrasting opinion on anything requires comparison. As it stands, there are no "golden standards" for evaluating these prompts, which is why we wanted to propose several readings for evaluating the effectiveness of a prompt, inspired by studies on absolute and relative performance (Kalloori et al., 2019; Roch et al., 2007; Roscoe et al., 1999). In this type of study, it is generally emphasized that there are several examples of a task to be performed (for us, the identical prompt) in the evaluation of situations or texts (Frisbie & Waltman, 1992; Gill & Bramley, 2013).

Tools like ChatGPT are able to facilitate this seamlessly, offering us several solutions and letting us choose which one we prefer to suit our style. However, this strategy does not provide viable answers immediately, which could be improved by the virtuous loop discussed earlier.

Therefore, we have conducted a pilot study with two conditions, both with the same general objective: to read one or more outputs from a generative AI and to evaluate them in two groups. For the first group, the aim was to carry out a set of systematic evaluations of the results of a prompt: reading the prompt, evaluating it on a scale ranging from 1 (very bad) to 5 (very good), then describing their reflections, clearly stating the positive and negative elements, and distinguishing avenues for improvement from the commentary. In the second group, the aim is to study several outputs that are still based on the same Gen-AI and the same prompt. This is in line with Shah's (2024) remarks on how to make prompt and outcome studies more scientific. In cases where concrete contextual feedback is required, users will be more able and motivated to provide it if they are presented with two outcomes from the same prompts and asked to rate and comment on both on an absolute scale, as opposed to the rule of displaying only one outcome to rating at a time?

We conducted an experiment with 32 Danish high-school students divided into two groups to examine the impact of varying prompts on their critical response skills. The prompt was designed to be relevant to problems that students may encounter in real life, and aimed to assess the feasibility of incorporating AI into their daily school activities. Based on the overall school-task, the prompt produced a detailed a task breakdown and for each made suggestions for the potential level and method of incorporating AI into the subtask solution giving a human-AI task hybridization level ranging from 1 (human performed the majority of the task) to 5 (AI performed the majority of the task). This prompt was used multiple times in different ChatGPT4 conversations to ensure the presence of two distinct AI individuals. The first group (NGr1 = 19) received only one response, while the second group (NGr2 = 13) received two, one identical to the first group and the other different. Both groups were asked to describe the outcome, rate it on a scale of 1 to 5 (1 being very bad and 5 being excellent) and provide justification for their scores.

Although our sample size was small, we found it interesting to examine the descriptive differences between the two groups. Most notably there was an inverse difference in the number of words used, with Group 1 using fewer words (mean of 21) than Group 2 (mean of 26), which might indeed suggest that Group 2 engaged more deliberately in the evaluation process. Because additional research is necessary to establish a definitive conclusion, we believe that these initial descriptive findings are worth exploring. It is possible that the number of words used is an indicator of the quality of a student's justification or argument (Abrami et al., 2008).

Despite the fact that our research did not yield statistically significant outcomes as a result of the limited number of participants, it is noteworthy that, at least in terms of descriptive statistics, disparities do exist and appear to support our hypothesis. Further investigation could be conducted by engaging larger sample sizes, examining participants' critical abilities, broadening the demographic scope, and considering additional variables.

Abrami, P. C., Bernard, R. M., Borokhovski, E., Wade, A., Surkes, M. A., Tamim, R., & Zhang, D. (2008). Instructional Interventions Affecting Critical Thinking Skills and Dispositions : A Stage 1 Meta-Analysis. Review of Educational Research, 78(4), 1102-1134. https://doi.org/10.3102/0034654308326084

Frisbie, D. A., & Waltman, K. K. (1992). Developing a Personal Grading Plan. Educational Measurement: Issues and Practice, 11(3), 35–42. https://doi.org/10.1111/j.1745-3992.1992.tb00251.x

Gill, T., & Bramley, T. (2013). How accurate are examiners' holistic judgements of script quality? Assessment in Education: Principles, Policy & Practice, 20(3), 308–324. https://doi.org/10.1080/0969594X.2013.779229

Kalloori, S., Li, T., & Ricci, F. (2019). Item Recommendation by Combining Relative and Absolute Feedback Data. Proceedings of the 42nd International ACM SIGIR Conference on Research and Development in Information Retrieval, 933–936. https://doi.org/10.1145/3331184.3331295

Roch, S. G., Sternburgh, A. M., & Caputo, P. M. (2007). Absolute vs Relative Performance Rating Formats: Implications for fairness and organizational justice. International Journal of Selection and Assessment, 15(3), 302–316. https://doi.org/10.1111/j.1468-2389.2007.00390.x

Roscoe, E. M., Iwata, B. A., & Kahng, S. (1999). Relative Versus Absolute Reinforcement Effects: Implications for Preference Assessments. Journal of Applied Behavior Analysis, 32(4), 479–493. https://doi.org/10.1901/jaba.1999.32-479

Shah, C. (2024). From Prompt Engineering to Prompt Science With Human in the Loop (arXiv:2401.04122). arXiv. https://doi.org/10.48550/arXiv.2401.04122Sharaf, Y.